\documentclass[prb,showpacs]{revtex4}

\usepackage{graphicx}
\usepackage{dcolumn}
\usepackage{bm}
\usepackage{amsmath}
\usepackage{amssymb}
\usepackage{latexsym}
\usepackage{epsfig}
\usepackage{amsbsy}
\usepackage{array}
\usepackage{amssymb}
\usepackage{setspace}
\usepackage{bm}
\usepackage{algorithm}

\def\sint{\ifmmode{- \!\!\!\!\!\! \int}
    \else{\hbox{$- \!\!\!\! \int \ $}}\fi}

\begin{document}

\title{Topological Phase Transition in Single Crystals of \bm{$(Cd_{1-x} Zn_x )_3 As_2$}}

\author{Hong Lu$^1$}
\author{Xiao Zhang$^1$}
\author{Shuang Jia$^{1,2,}$\footnote{Author to whom correspondence should be
		addressed; electronic mail: gwljiashuang@pku.edu.cn.}}
\affiliation{$^1$ICQM, School of Physics, Peking University, Beijing 100871, China}
\affiliation{$^2$Collaborative Innovation Center of Quantum Matter, Beijing, China}
             
\begin{abstract}
	Single crystals of $(Cd_{1-x} Zn_x)_3 As_2$ were synthesized from high-temperature solutions and characterized in terms of their structural and electrical properties. Based on the measurements of resistivity and Hall signals, we revealed a chemical-doping-controlled transition from a three dimensional Dirac semimetal to a semiconductor with a critical point $x_c\sim 0.38$. We observed structural transitions from a body-center tetragonal phase to a primary tetragonal phase then back to a body-center tetragonal phase in the solid solutions as well, which are irrelevant to the topological phase transition. This continuously tunable system controlled by chemical doping provides a platform for investigating the topological quantum phase transition of 3D Dirac electrons. 
\end{abstract}

\pacs{06.60.-c, 61.05.-a, 71.30.+h, 72.90.+y}

\maketitle

\section{Introduction}
Cadmium and zinc pnictides ($Cd_3 As_2$, $Zn_3 As_2$, $Cd_3 P_2$ and $Zn_3 P_2$) belong to a group of $A_3^{\uppercase\expandafter{\romannumeral2}} B_2^{\uppercase\expandafter{\romannumeral5}} $ semiconductors and semimetals, which have been well known for their potential applications in high efficiency solar cells and optoelectronic devices \cite{reff1,reff2,reff3,reff4}. These four compounds crystallize at various temperatures in several closely related structures, which can be viewed as the different arrangements of a distorted antifluorite structure \cite{reff5,reff6,reff7,reff8,reff9,reff10}.

The electrical properties of these four compounds are distinct in several aspects. $Zn_3 As_2$,  $Zn_3 P_2$ and  $Cd_3 P_2$ are semiconductors with low carrier mobility and direct band gaps being 1.0 eV, 1.5 eV and 0.5 eV respectively \cite{reff11,reff12,reff13}. Both $Zn_3 P_2$ and  $Zn_3 As_2$ are $p$-type, while  $Cd_3 P_2$ and  $Cd_3 As_2$ are $n$-type \cite{reff11,reff12,reff13}. On the other hand, previous studies on the optical properties of $Cd_3 As_2$  suggested that it was a semiconductor with a narrow band gap around  $0.1 eV$ \cite{reff11}. The mobility for  $Cd_3 As_2$  was reported as high as $1.5\times 10^4\ cm^2/Vs$ at room temperature \cite{reff11}. For comparing, the hole mobility for $Zn_3 As_2$ was only $10 \ cm^2/V s$ at room temperature. $Cd_3 As_2$ was believed to manifest an inverted band structure due to the spin-orbital coupling (SOC) while the other three had normal band structures \cite{reff14}.

Very recent studies on $Cd_3 As_2$ have revealed the topological aspect of its electrical properties \cite{reff15,reff16,reff17,reff18,reff19,reff20,reff21,reff22,reff23}. Band structure calculation predicted that  $Cd_3 As_2$ was a three dimensional (3D) Dirac semimetal with the band inversion \cite{reff15}. The energy dispersion of the Dirac electron is protected by the rotational symmetry along the crystallographic $\bf c$ axis in the tetragonal unit cell. The 3D Dirac cones of $Cd_3 As_2$ have been observed in angle-resolved photoemission spectroscopy (ARPES) \cite{reff16,reff17,reff18}. The Dirac-like band dispersion and the inverted band ordering was probed by the Landau level spectroscopy and quasiparticle interference in scanning tunneling microscopy (STM) \cite{reff19}. Based on the electrical transport measurements, two experimental groups found an ultrahigh mobility of the Dirac electrons \cite{reff20,reff21}. A strongly sample-dependent, large linear magnetoresistance ($MR$)  was observed in $Cd_3 As_2$ at low temperatures \cite{reff20}. The nonsaturating linear $MR$ in $n$-type $Cd_3 As_2$ up to $65T$ was believed to result from its mobility fluctuations \cite{reff23}. The anisotropic Fermi surface with two ellipsoids of Dirac electrons along the $\bf c$ axis was revealed from the angular dependent measurements of SdH oscillations \cite{reff21}.

Noticing the opposite band orderings in $Cd_3 As_2$ and the other three members in the family, we expected a band inversion transition in pseudo-binary compounds of $(Cd_{1-x} Zn_x)_3 As_2$ and ${Cd_3 (As_{1-x} P_x )_2}$. A band inversion transition due to the change of the SOC strength has been observed in solid solutions of semiconductors and semimetals such as ${Hg_{1-x} Cd_x Te}$ and ${Pb_{1-x} Sn_x Se}$ \cite{reff24,reff25}. Recent studies on the solid solutions of ${TlBiSe_{2-x} S_x}$  and ${Bi_{2-x} In_x Se_3}$ have confirmed the existence of a quantum phase transition tuned by chemical doping from topological insulators to trivial band insulators \cite{reff26,reff27}. A topological phase transition from a 3D Dirac semimetal to a trivial semiconductor was predicted in ${Na_3 Bi_{1-x} Sb_x}$ and ${Cd_3 (As_{1-x} P_x )_2}$ by first-principle calculations as well \cite{reff28}.

The changes of the structures and physical properties of polycrystalline ${(Cd_{1-x} Zn_x)_3 As_2}$ and ${Cd_3 (As_{1-x} P_x )_2}$ have been studied \cite{reff29,reff30,reff31,reff32}. ${(Cd_{1-x} Zn_x)_3 As_2}$ crystallize in a primary tetragonal structure \cite{reff29}. The majorities undergo a crossover from $n$- to $p$-type when $x$ increases in $(Cd_{1-x} Zn_x)_3 As_2$ \cite{reff30,reff31}, while the band gap increases linearly with the proportion of $Zn$, according to the magneto-optical measurements \cite{reff32}. Due to the lack of single crystals, the change of the band structure and the electrons, especially in the topological aspect, cannot be addressed in detail.

In this study, we report the single-crystalline $(Cd_{1-x} Zn_x )_3 As_2$ obtained from high-temperature solution growth. Powder X-ray Diffraction (XRD) measurements revealed structural transitions from a body-center tetragonal phase for  $Cd_3 As_2$, to a primary tetragonal phase for $0.07\leq x<0.52$, and then back to a body-center tetragonal phase for $x>0.52$. The electrical resistivity and Hall measurements revealed a metal-insulator transition at a critical point $x_c\sim 0.38$. The analysis of the $MR$ demonstrated a transition from a 3D Dirac semimetal to a trivial direct-gap semiconductor via the modulation of the SOC strength.

\section{experiment}
Single crystalline $(Cd_{1-x} Zn_x )_3 As_2$ samples were grown from high temperature solutions with the initial concentration of starting elements being  $(Cd_{1-y} Zn_y )_9 As_1$. The mixtures were sealed in evacuated quartz ampoules, and then kept at a high temperature between $800^{\circ}C$ and $1100^{\circ}C$ for two days, and then slowly cooled down to $425^{\circ}C$ with a rate of $-5^{\circ}C/$hour. After staying at $425^{\circ}C$  for one day, the ampoules were centrifuged to separate crystals from flux. The single crystals of $Cd_3 As_2$ were mainly 3D bulks with triangular facets, but several needle-like crystals were found in the growth as well. The shapes of the crystals were similar as that described in Ref.$\ $[20,21]. When zinc was added to the solutions, the sizes of the 3D crystals decreased, while some flake-like crystals with smooth or mesa-landscape-like surfaces appeared. For $x\geq 0.58$, the flake-like crystals were dominant and no 3D crystals appeared in the growth. XRD measurements revealed that both the triangular facets (Fig. 1(f)) and the large surface of the flake-like crystals (Fig. 1(e)) were either the $(011)_P$ face of the $P4_2/nmc$ structure or the $(112)_I$ face of the $I4_1/cd$ structure, which were the counterparts of each other (Fig. 2(c)-2(d)). In a same batch of growth, the crystals with different morphologies did not show larger difference of physical properties than those with the same morphologies.

In order to determine the zinc concentration $x$, we measured the Energy Dispersive X-ray Spectrum (EDX) of the samples in an FEI Nova NanoSEM 430 spectrometer. The samples with no residual cadmium were selected in the measurements and their EDX spectrum was observed through an overall area scanning. The linear relation of the measured zinc concentrations $x$ and the initial $y\ (0\leq y\leq 0.1)$ is shown in Fig. 3(c). By weighing the mass of the crystals yielded in every growth, we found that all the initial stoichiometric zinc was compounded in the crystals.

The powder XRD data was collected from a Rigaku MiniFlex 600 diffractometer and then refined by a Rietica Rietveld program. As shown in Fig. 3(b), the lattice constants and the volume of the unit cell change linearly with $x$ in accordance with a Vegard’s law. This result is same with what observed in polycrystalline $(Cd_{1-x} Zn_x)_3 As_2$ \cite{reff33}. Therefore we selected $x$ determined by EDX measurements as the nominal zinc concentrations, which were estimated to have less than $\pm 1\%$ difference between the real zinc concentrations. More details of the XRD experiments are discussed in the Result part.

Single crystals were polished to bars with length $\sim 1.0mm$, width $\sim 0.4mm$ and thickness $\sim 0.3mm$ for electrical transport measurements. The crystallographic orientation of the bars was same as those chosen in the recent experiments \cite{reff18,reff19,reff20,reff21,reff22}, i.e., the current was perpendicular to the $[112]_I$ direction in $I4_1/cd$ (or $[011]_P$ direction in $P4_2/nmc$), and the magnetic field was along the $[112]_I$ direction in $I4_1/cd$ (or $[011]_P$ direction in $P4_2/nmc$). The electrical resistance and Hall voltage were measured via a four-point contact method in Quantum Design Physical Property Measurement System (PPMS-9).

\section{result}
The low temperature phases of $Cd_3 As_2$ and $Zn_3 As_2$ were reported as $\alpha''$ ($P4_2/nmc$), $\alpha'$ ($P4_2/nbc$) and  $\alpha$ ($I4_1cd$) at different temperatures \cite{reff5,reff6,reff7}, which evolve from a high-temperature  $\beta$ ($Fm\bar{3}m$) phase \cite{reff6,reff8}. The $\beta$ phase belongs to an antifluorite structure in which an arsenic atom is coordinated by six cationic and two vacancies randomly distributed in corners of a cube. In the low-temperature phases, two cationic atoms are missed along a diagonal of one face in the distorted cube \cite{reff5,reff6,reff7}. At room temperature, both $Cd_3 As_2$ and $Zn_3 As_2$ were reported to crystallize in a body-center tetragonal phase \cite{reff6,reff7,reff16,reff17,reff18,reff19,reff20,reff21,reff22,reff23}. Recently single-crystal XRD measurements suggested that the crystals of $Cd_3 As_2$ form in the structure of $I4_1/acd$ instead of $I4_1cd$  at room temperature \cite{reff9}. The unit cell of $I4_1cd$ phase is made of the unit cells of $P4_2/nmc$ phase associated with the lattice constants $a_I=\sqrt{2}a_P$ and $c_I= 2c_P$ (The subscripts $I$ and $P$ present the body-center and primary space group respectively).

Our powder XRD measurements revealed that $(Cd_{1-x} Zn_x )_3 As_2$ have different crystal structures for different $x$ at room temperature. Figure 2(a) shows that the crystals of $Cd_3 As_2$ grown from flux are $\alpha$ phase, and their XRD pattern exhibits the characteristic peaks of $(231)_I$, $(233)_I$ and $(237)_I$ of the body-center tetragonal phase for $25^\circ<2\theta<36^\circ$. This result is consistent with what was previously reported \cite{reff7,reff16}. Limited by the quality of the data, we could not determine whether $Cd_3 As_2$ crystallizes in the space group of $I4_1cd$ or $I4_1/acd$. Once a small amount of $Zn$ was added ($x=0.07$), the XRD pattern is distinct from that of $Cd_3 As_2$. The characteristic peaks of $(231)_I$, $(233)_I$ and $(237)_I$ disappear, while the $(032)_P$ peak of the $P4_2/nmc$ group occurs (Fig. 2(a)). This peak remains resolvable until the doping level reaches $x=0.38$. For $0.38<x<0.46$, the structure reenters the body center tetragonal structure $I4_1cd$ accompanied by $(240)_I$ and $(244)_I$ peaks which are exceedingly weak in the pattern of $Cd_3 As_2$ \cite{reff7,reff34}(Fig. 2(b)).

The peaks of $(440)_I$ with the strongest intensity stand at $40.0^\circ$ and $43.1^\circ$ for $\alpha-Cd_3 As_2$ and  $Zn_3 As_2$ respectively (Fig. 3(a)). The peak of $(040)_P$ is the counterpart of $(440)_I$ in $P4_2/nmc$ group. Figure 3(a) shows that the peaks of $(040)_P$ and $(440)_I$ shift gradually when $x$ changes from $0$ to $1$.  Although the volume of the unit cell changes about $20\%$ from $x=0$ to $1$, the peak shape does not change significantly in the solid solutions, indicating homogeneous chemical distributions in the crystals. The lattice constants for the samples in the $I4_1cd$ group were presented in the view of the $P4_2/nmc$ group as $a_I=\sqrt{2}a_P$ and $c_I=2c_P$. Figure 3(b) shows that $a_P$ and $c_P$ change in a precisely linear relation with respect to $x$, albeit the structural transitions.

Temperature dependent resistivity of $(Cd_{1-x} Zn_x)_3 As_2$ shows a clear change from a metallic to semiconducting profile when $x$ increases from $0\sim 0.31$ to $0.38\sim 0.58$ (Fig. 4(a)). The resistivity of $Cd_3 As_2$ is close to what was reported in Ref.$\ $[22] with the residual resistivity ratio ($RRR=\rho_{(T=300K)}/\rho_{(T=2K)}$) being 10 (Fig. 8).  The values of  $RRR$ keep almost invariant when $x$ increases up to $0.31$ (Fig. 8). For $x=0.38$, the $\rho(T)$ decreases with decreasing temperatures above $200 K$, and then increases below this temperature. This complicated behavior indicates that the sample is likely a very narrow bandgap semiconductor. The values of $RRR$ then dramatically decrease for $x\geq 0.38$, being $0.58$ for $x=0.38$ and $7.7\times10^{-6}$ for $x=0.58$ (see more details in Fig. 8). The changes of the $RRR$ for different $x$ indicate a process of band gap opening for $x\geq 0.38$.

Figure 4(b) shows the $MR$ for the samples for $x\leq 0.46$ at $2K$. When $x\leq 0.31$, the values of the $MR$ are comparably large as that for $Cd_3 As_2$ \cite{reff22,reff35}. The values of the $MR$ decline significantly for the semiconducting samples for $x\geq 0.38$. Recent studies of $Cd_3 As_2$ reported linear $MR$ at low temperatures \cite{reff22,reff23}. In our experiments, the $MR$ of $(Cd_{1-x} Zn_x )_3 As_2$ follows the power law of $MR\propto H^{\alpha}$ where $\alpha$ varies from $0.9$ to $1.5$ for different samples.

Conspicuous SdH oscillations occur in the field dependent resistivity for all the samples for $x\leq 0.38$ at low temperatures, while the oscillations were not observed for the samples for $x\geq 0.46$. This observation is consistent with a metal-insulator transition with a critical point $x_c\sim 0.38$. The part of resistivity with oscillations versus the reciprocal of the magnetic field is presented in Fig. 5. For $x<0.29$, only one frequency was observed for each sample (Fig. 6(a)). The frequencies of the oscillations show a clear trend of a decline with respect to $x$ up to $0.29$ (Fig. 6(a)). For $0.29\leq x\leq 0.38$, the frequencies show more significant sample difference in a same batch. Some samples show single frequencies from $15T$ to $30T$, while the second and third frequencies as large as $70T$ occur in other samples. Such strong sample-dependence and complicated multi-frequency features indicate that the samples for $0.29\leq x\leq 0.38$ are semimetals or very narrow bandgap semiconductors with complicated Fermi surface which is strongly influenced by subtle changes of chemical potential.

As shown in Fig. 5, the temperature dependent amplitudes of the SdH oscillations were fitted by the Lifshitz-Kosevich formula \cite{reff36,reff37,reff38}:

\[\Delta \rho_{xx} \propto A(T) e^{-\frac{2\pi^2k_BT_D}{\hbar \omega_c}}\cos 2\pi (\frac{S_F}{B}+\beta)\]

\[A(T)=\frac{2\pi^2 k_BT/\hbar \omega_c}{sinh(2\pi^2 k_BT/\hbar \omega_c)}\]

where $k_B$ is the Boltzmann's constant; $\omega_c$ is the cyclotron frequency; $T_D$ is the Dingle temperature and $A(T)$ is the thermal damping factor which helps to fit the energy gap $\hbar\omega_c$. For the samples with multi-frequencies, their main frequencies were analyzed. For large $x$, the amplitudes of the oscillations are damped less significantly by the temperatures for large $x$, which indicates a smaller cyclotron effective mass (Fig. 5).

The parameters of the SdH oscillations for different $x$ are listed in Table $\uppercase\expandafter{\romannumeral1}$. The cross-sectional area $A_F$ in the momentum space comes from the Onsager relation $S_F=\frac{\hbar}{2\pi e}A_F$. Simply assuming a circular $A_F$, we got the Fermi wave vector $k_F$ from  $A_F=\pi k_F^2$. The Fermi velocity  $\nu_F=\hbar k_F/m^*$, the Fermi energy $E_F=\nu_F^2 m^*$, and the cyclotron effective mass $m^*=eB/\omega_c$  are listed as well. Figure 6 shows that $m^*$ changes in a similar manner as $S_{F1}$ with respect to $x$.

In order to better understand the metal-insulator transition, we measured Hall resistivity in $(Cd_{1-x} Zn_x )_3 As_2$ at $2K$. The field dependent Hall resistivity of the samples for $x\leq 0.31$ shows a linear negative profile with SdH oscillations on the background. The negative linear-field-dependent $\rho_{yx}(H)$ indicates that the carriers in the samples for $x\leq 0.31$ simply originate from an electron band. The carrier density decreases linearly with increasing $x$ from $0$ to $0.31$ (Fig. 7(a)). These results are consistent with the observation of the decreasing SdH oscillation frequencies with respect to $x$. $\rho_{yx}(H)$ becomes smaller and nonlinear for $0.38\leq x<0.59$. This nonlinear feature is clear for $x=0.46$ (inset of Fig. 7(a)). In this range, the samples manifest a semiconducting $\rho(T)$ profile while multi-frequencies were observed in the SdH oscillations in their $MR$. The Hall signals and the resistivity indicate two types of carriers. For $x\geq 0.59$, the Hall signals become large and positively field-dependent, which indicate $p$-type semiconductors consistent with low carrier concentrations. The change of the carrier density $n_H$ and mobility $\mu_c$ with respect to $x$ is summarized in Fig. 7(b).

\section{discussion}
Our measurements show that the change of the electrical properties of  $(Cd_{1-x} Zn_x )_3 As_2$ has no observable correlation with the structural transitions. This result is not unexpected according to previous band structural calculation. Both $\alpha''$ $(P4_2/nmc)$ and  $\alpha$ $(I4_1cd)$ phases of $Cd_3 As_2$ manifest similar simple band structures near the Fermi surface. Only two Dirac cones protected by rotational symmetry cross the Fermi level $(E_F)$ along the high symmetric line $\Gamma-Z$ in the Brillouin zones \cite{reff15}. Therefore a structural transition cannot influence the bands near the $E_F$.

Previous studies showed that the calculation and experiments, the negative gap is about $-0.3eV\sim -0.7eV$ for $Cd_3 As_2$ \cite{reff15,reff32}, while the direct gap is $1.0eV$ for $Zn_3 As_2$ \cite{reff11}. With a semimetal and a semiconductor as two terminals, the band inversion transition should accompany a metal-semiconductor transition at a certain $x$. If we assume that the band gap of $(Cd_{1-x} Zn_x )_3 As_2$ changes linearly with respect to $x$ \cite{reff32}, the critical point of the band inversion transition is estimated to occur in the range of $0.23\leq x\leq 0.41$. This estimation is consistent with our experimental results. The critical point can also be estimated by considering the change of the SOC strength in $(Cd_{1-x} Zn_x )_3 As_2$. Here we assume that the band inversion is solely induced by the change of the SOC strength, which is proportional to $Z^{4}/n^3$ in case of the hydrogenic wavefunctions in a Coulomb field where $Z$ is the nuclear charge and $n$ is the principal quantum number \cite{reff39}. Then the critical point is estimated as $x\sim 0.35$, which is very close to the experimental result: $x_c\sim 0.38$.

The band structure calculation for $Cd_3 (As_{1-x} P_x )_2$ revealed a topological phase transition from a Dirac semimetal to a trivial semiconductor induced by the change of the SOC strength \cite{reff28}. When $x$ increases, the two Dirac points along the $k_z$ axis gradually move closer and then merge at the $\Gamma$ point under the protection of the crystal symmetry. A direct band gap is opened beyond the critical point. The process of the band inversion transition in $(Cd_{1-x} Zn_x )_3 As_2$ should be similar as that in $Cd_3 (As_{1-x} P_x )_2$. Despite of large chemical replacement in the crystals, the Dirac cones are robust, which is supported by the vanishing disorder self-energy around the crossing points \cite{reff40}. Further investigation such as ARPES measurements for $(Cd_{1-x} Zn_x )_3 As_2$ will help to reveal the details of this topological phase transition.

$Cd_3 As_2$ is always $n$-type due to $As$ vacancies, while $Zn_3 As_2$ is $p$-type because extra $Zn$ vacancies serve as electron acceptors \cite{reff11,reff35}. Since both two types of carriers come from element vacancies, an $n$ to $p$ transition is expected in $(Cd_{1-x} Zn_x )_3 As_2$. With increasing $x$, the zinc doping will suppress the chemical potential, which crosses a small Fermi surface near the Dirac cones. The decrease of $S_{F1}$ with increasing $x$ is a comprehensive result of the change of the band structure and chemical potential.

For $0.29\leq x\leq 0.38$, we found strongly sample-dependent frequencies of SdH oscillations. For a very narrow bandgap semiconductor or semimetal, any small change of the carrier concentrations will affect the chemical potential dramatically near the band touching. The strong sample-dependence and the complicated SdH oscillations in this regime are not unexpected.

\section{Summury}
Single crystals of  $(Cd_{1-x} Zn_x )_3 As_2$ were synthesized from high-temperature solutions. Based on the analysis of the electrical properties, we realized a transition from a 3D Dirac semimetal to a semiconductor with the critical point $x_c\sim 0.38$ in these solid solutions. The structural transitions do not affect the electrical properties in this system. The topological aspect of this metal-insulator transition needs experimental exploration in the future.

\section{Acknowledgements}
The authors acknowledge the members in Jian Wang's group and Yuan Li's group for helpful discussions and using their instruments. Chenglong Zhang, Zhujun Yuan and Cheng Guo in Jia's group are to be thanked heartily sincerely. The project is supported by National Basic Research Program of China (Grant Nos.2013CB921901 and 2014CB239302).

\newpage
\begin{figure}
\centering
\includegraphics[width=6in]{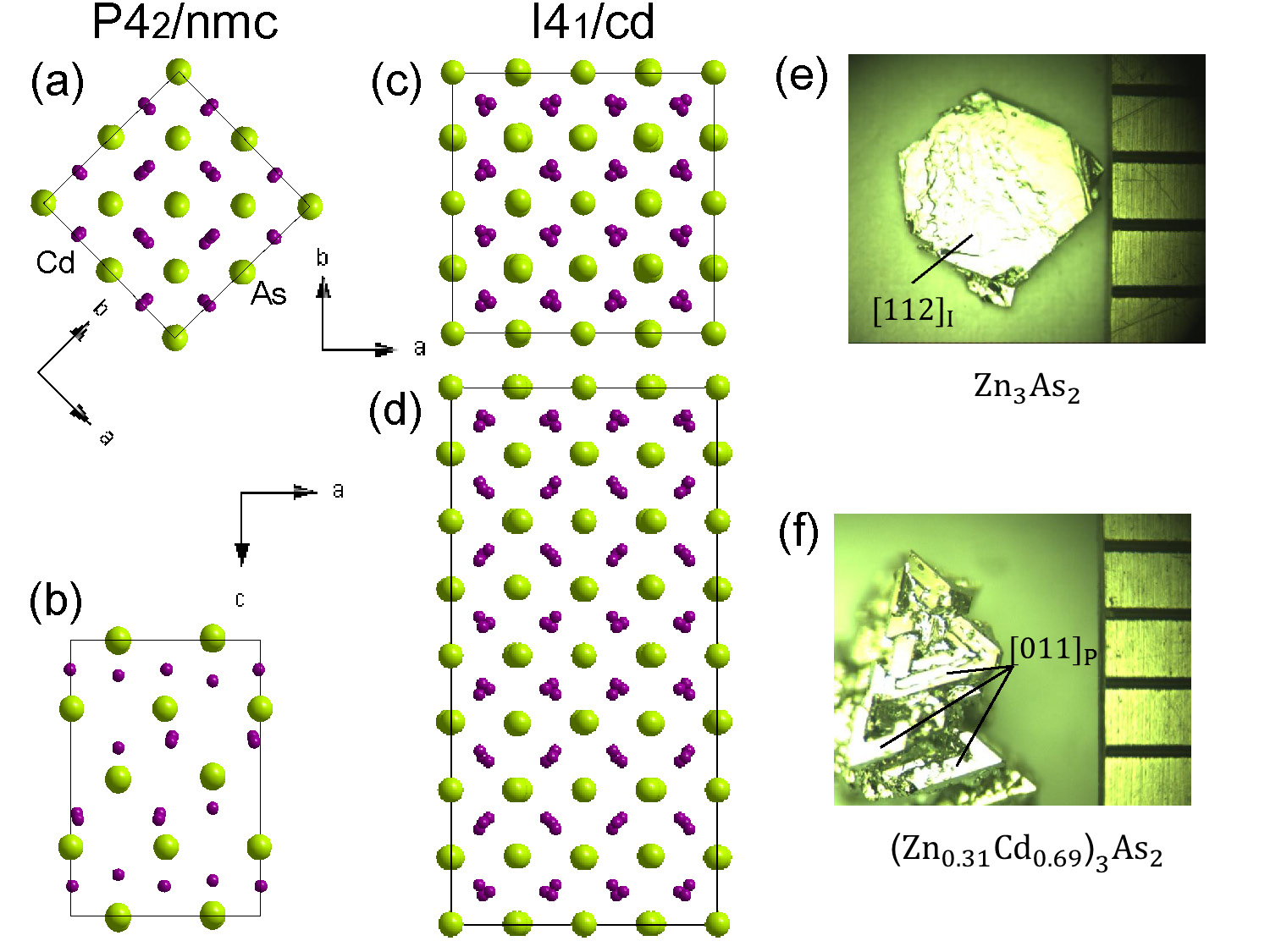}
\caption{Panel (a)-(d): Unit cells of $\alpha''-Cd_3 As_2 $ and $\alpha-Cd_3 As_2$ viewed along $\bf c$ and $\bf b$ axes respectively. Panel (e) and (f): Single crystals of $Zn_3 As_2$ and  $(Zn_{0.31} Cd_{0.69})_3 As_2$ show different morphologies (see more details in the text). The scale on right is $1mm$. \label{1}}
\end{figure}
\clearpage

\newpage
\begin{figure}
\centering
\includegraphics[width=6in]{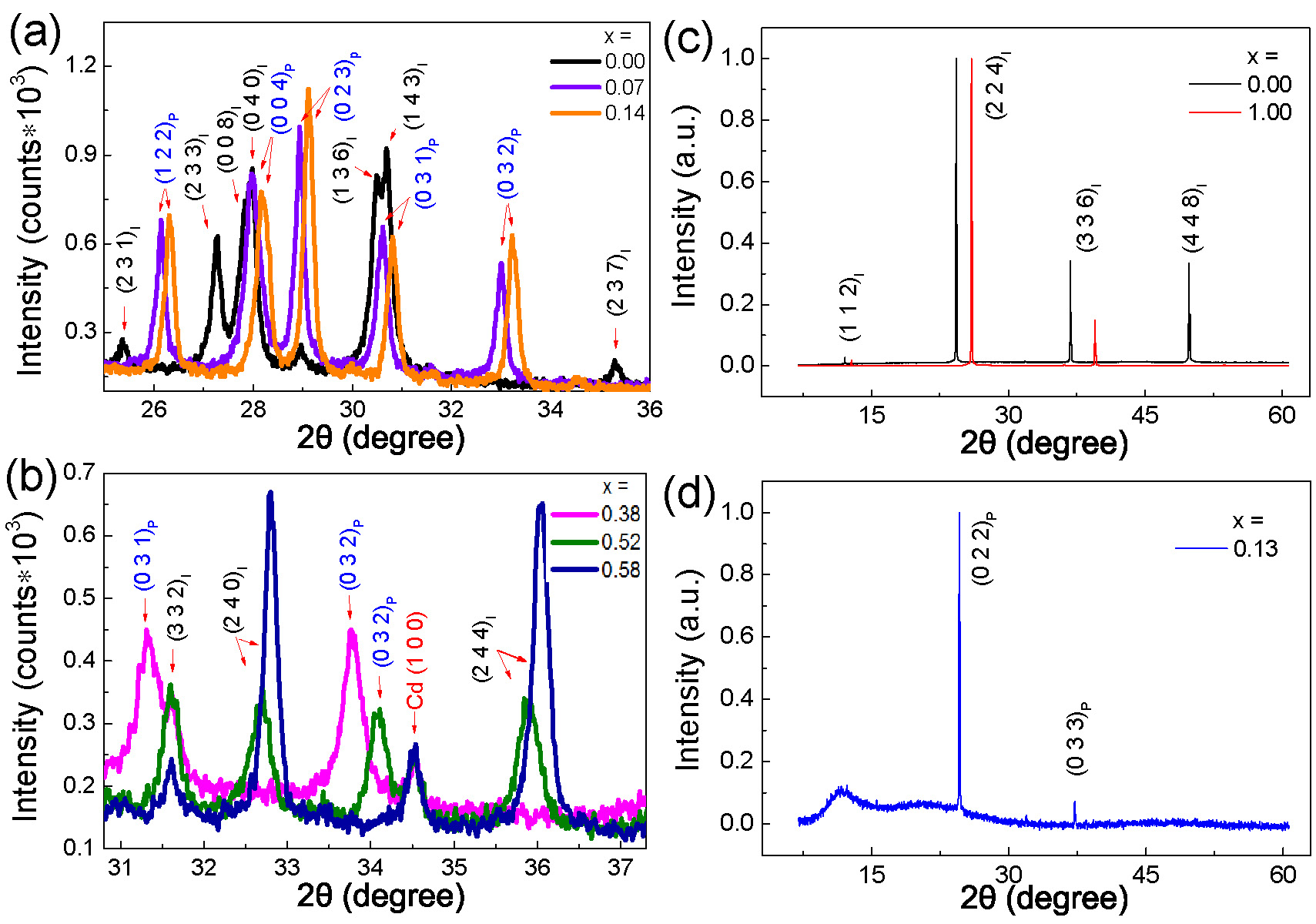}
\caption{\textbf{XRD patterns for \bm{$(Cd_{1-x} Zn_x)_3 As_2$}.} Panel (a): Powder XRD patterns for $x=0, 0.07$ and $0.14$ for $25^\circ<2\theta<36^\circ$. The characteristic peaks of $(231)_I$, $(233)_I$ and $(237)_I$ for a body-center structure and $(032)_P$ for a primary structure are labeled. Panel (b): Powder XRD patterns for $x=0.38, 0.52$ and $0.58$ for $31^\circ<2\theta<37^\circ$. The peak of $(032)_P$ occurs for $x=0.38$, while the peaks of $(240)_I$ and $(244)_I$ occur for $x=0.58$. For $x=0.52$, we observed two sets of peaks, which indicates the batch has the mixture of two types of crystals. The $(100)$ peak of cadmium at $34.5^\circ$ occurs in all the XRD patterns with no shift.  Panel (c): The diffraction pattern of the $(112)_I$ plane for $\alpha-Cd_3 As_2$ and  $\alpha-Zn_3 As_2$. Panel (d): The diffraction pattern of the $(022)_P$ plane for $(Cd_{0.87} Zn_{0.13})_3 As_2$. \label{11}}
\end{figure}
\clearpage
\newpage

\begin{figure}
\centering
\includegraphics[width=4in]{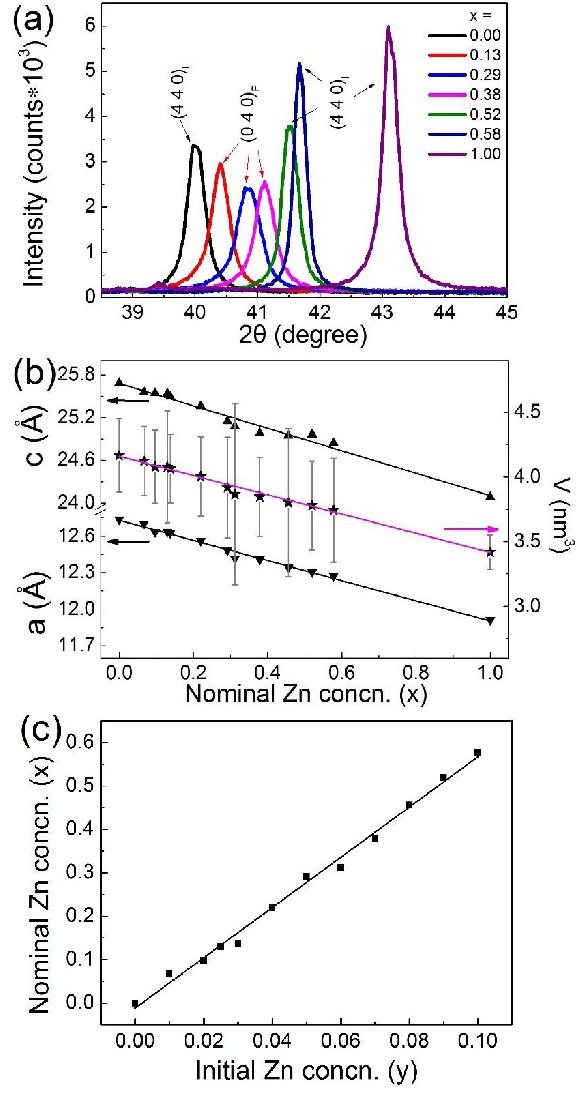}
\caption{Panel (a): The $(440)_I$ peaks and the counterpart $(040)_P$ peaks in the powder XRD patterns for $(Cd_{1-x} Zn_x)_3 As_2$. Panel (b): The lattice constants and the volume of the primitive cell change with the concentration (concn.) of zinc $x$ linearly. Panel (c): The nominal concentration  of zinc $x$ and the initial $y$ have a linear dependence for $0.00\leq x<0.58$. \label{}}
\end{figure}
\clearpage

\newpage
\begin{figure}
\centering
\includegraphics[width=4in]{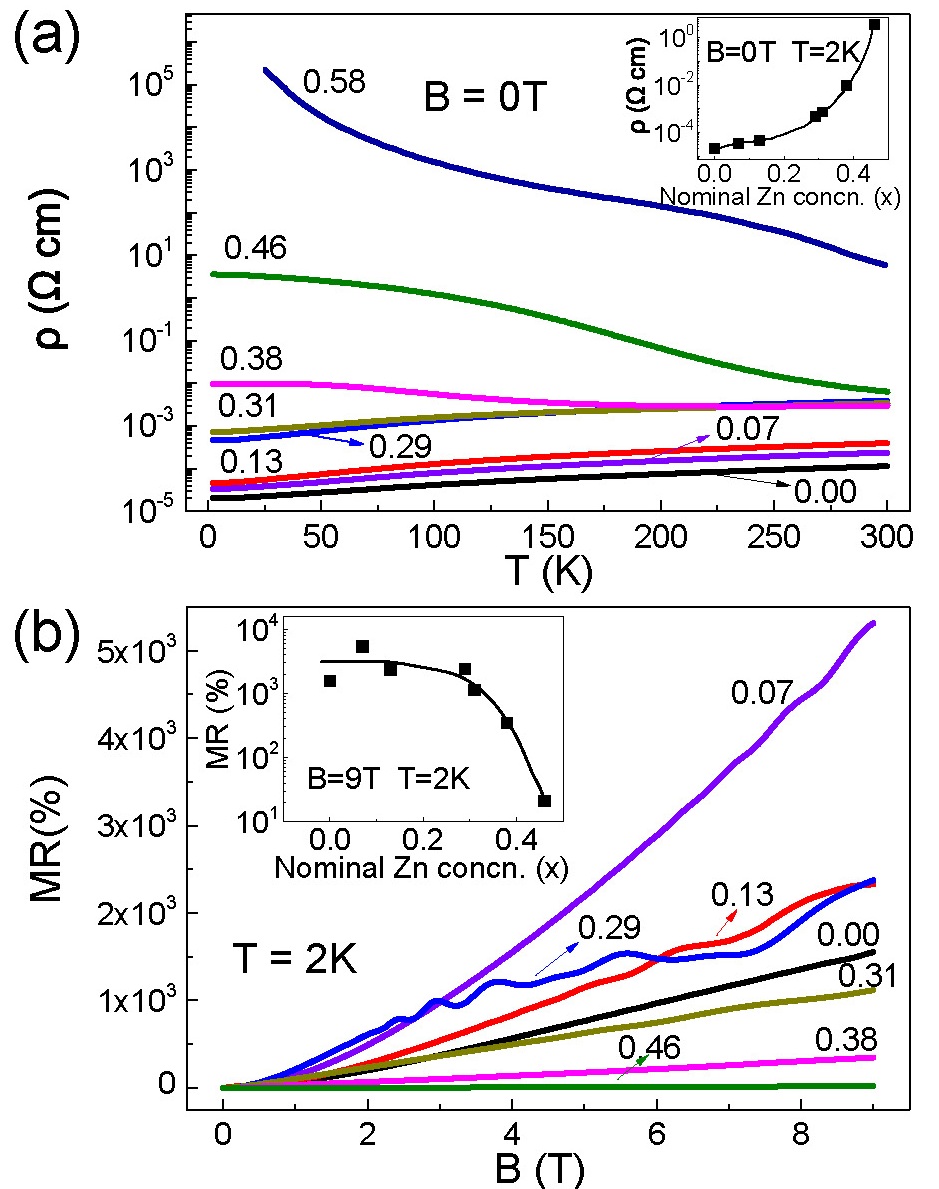}
\caption{Panel (a): Temperature dependent resistivity of $(Cd_{1-x} Zn_x)_3 As_2$ for $0\leq x\leq 0.58$ in zero magnetic field. Inset: The values of resistivity for $0 \leq x \leq 0.46$ at $2K$ in zero magnetic field. Panel (b): $MR$ versus magnetic fields at $2K$. Inset: The values of MR for $(Cd_{1-x} Zn_x)_3 As_2$ ($0\leq x \leq 0.58$) versus $x$ at $9T$ at $2K$.}
\end{figure}

\clearpage

\newpage
\begin{figure}
	\centering
	\includegraphics[width=4in]{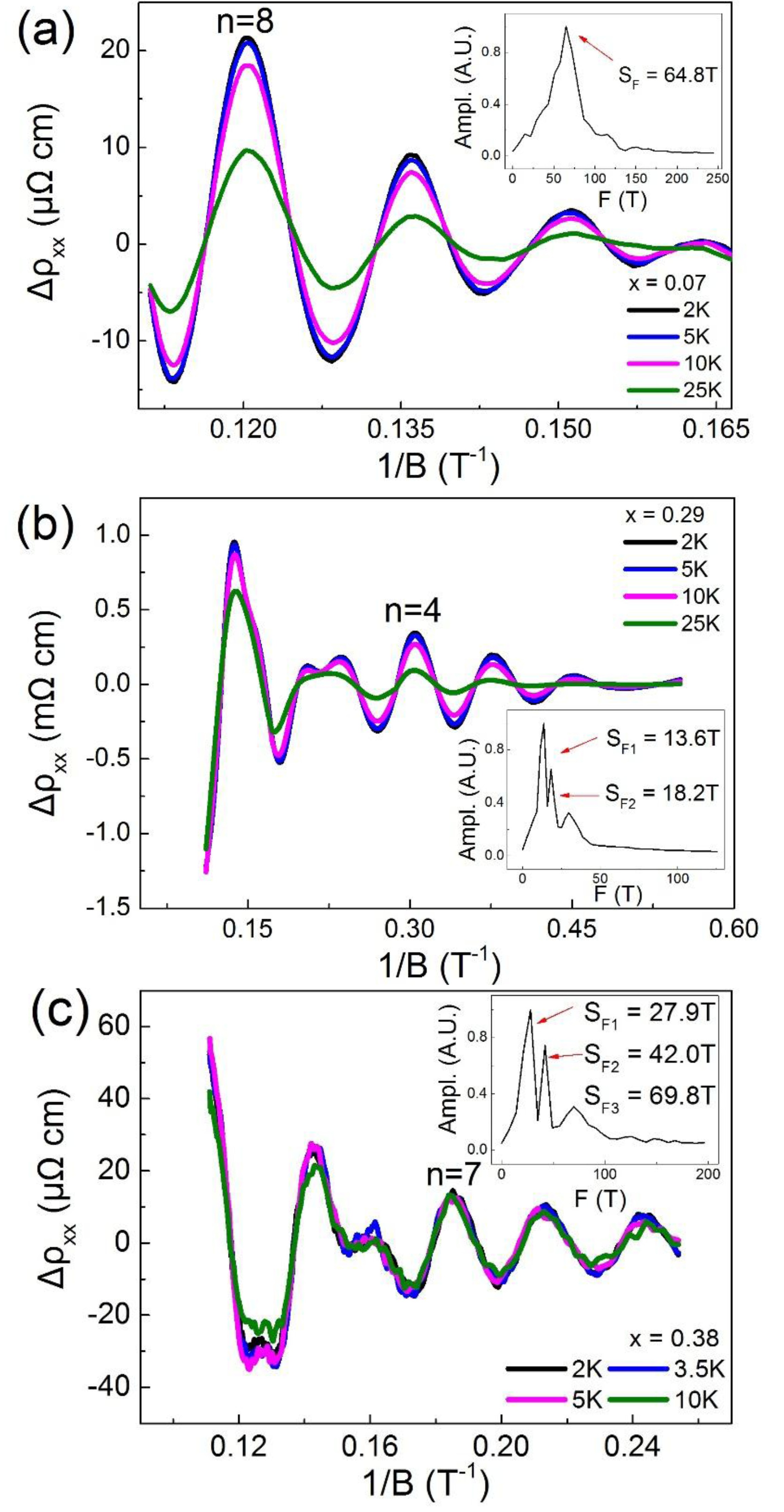}
	\caption{The oscillatory components of $\Delta\rho_{xx}$ versus the reciprocal of the magnetic field $(1/B)$ at different temperatures for $x=0.07, 0.29$ and $0.38$. Insets: Fast Fourier Transform (FFT) spectra for $x=0.07, 0.29$ and $0.38$.  The results of the SdH oscillations are summarized in Table $\uppercase\expandafter{\romannumeral1}$.}
\end{figure}

\clearpage

\newpage
\begin{figure}
	\centering
	\includegraphics[width=4in]{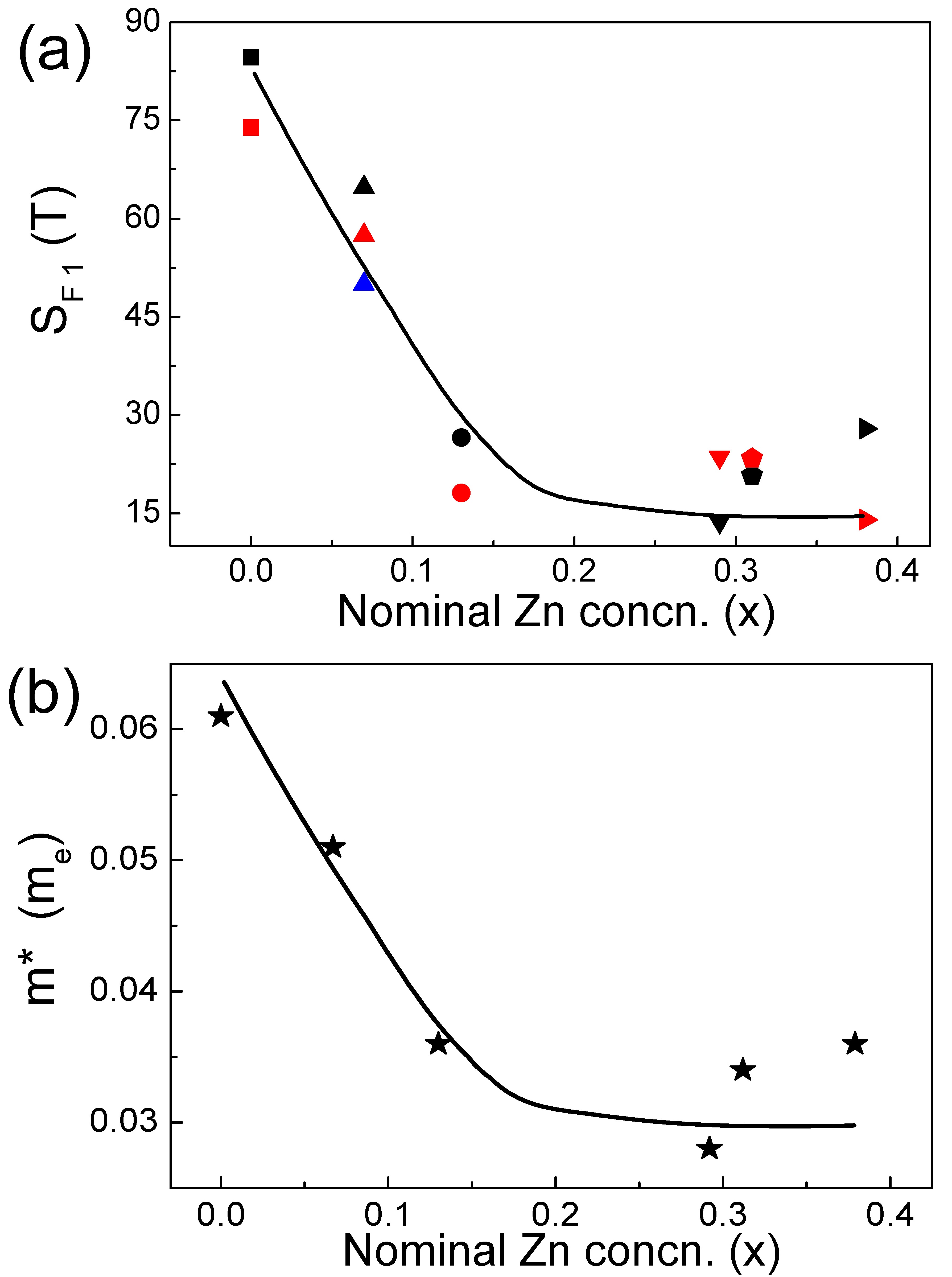}
	\caption{Panel (a): The main frequency of the SdH oscillations at $2K$ for $(Cd_{1-x} Zn_x)_3 As_2$. Different colors represent different samples from the same batches. The black ones are the samples corresponding to Fig. 4(a) and 7(b). Panel (b): The cyclotron effective mass $m^*$ according to the main frequency for each sample versus $x$. The line is for visual guidance.}
\end{figure}

\clearpage

\newpage
\begin{figure}
	\centering
	\includegraphics[width=4in]{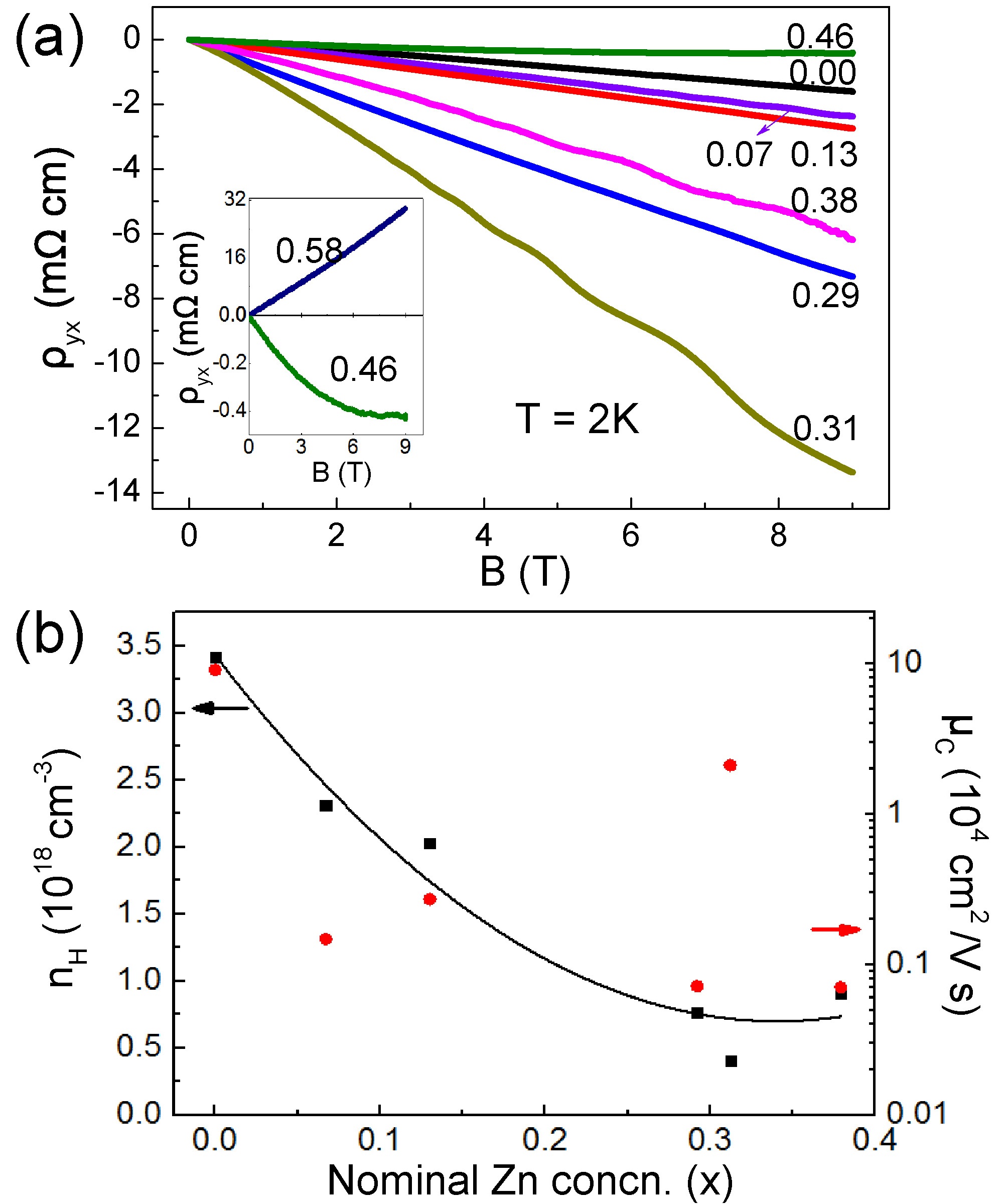}
	\caption{Panel (a): Field dependent Hall resistivity of $(Cd_{1-x} Zn_x)_3 As_2$ for $0\leq x \leq 0.46$ at $2K$. From $x=0.00$ to $0.46$, all samples are $n$-type. Inset: The Hall resistivity of the sample for $x=0.46$ shows a non-linear profile. For $x=0.58$, the signal turns to $p$-type. Panel (b): Carrier density $n_H$ $(n_H=B/(e\rho_{yx}))$ and mobility $\mu_c(\mu_c={1}/{(e \rho_{xx} n_H)})$ at $2K$ versus $x$ for $x=0.00$ to $0.38$. The solid line is for visual guidance.}
\end{figure}

\clearpage

\newpage
\begin{figure}
	\centering
	\includegraphics[width=4in]{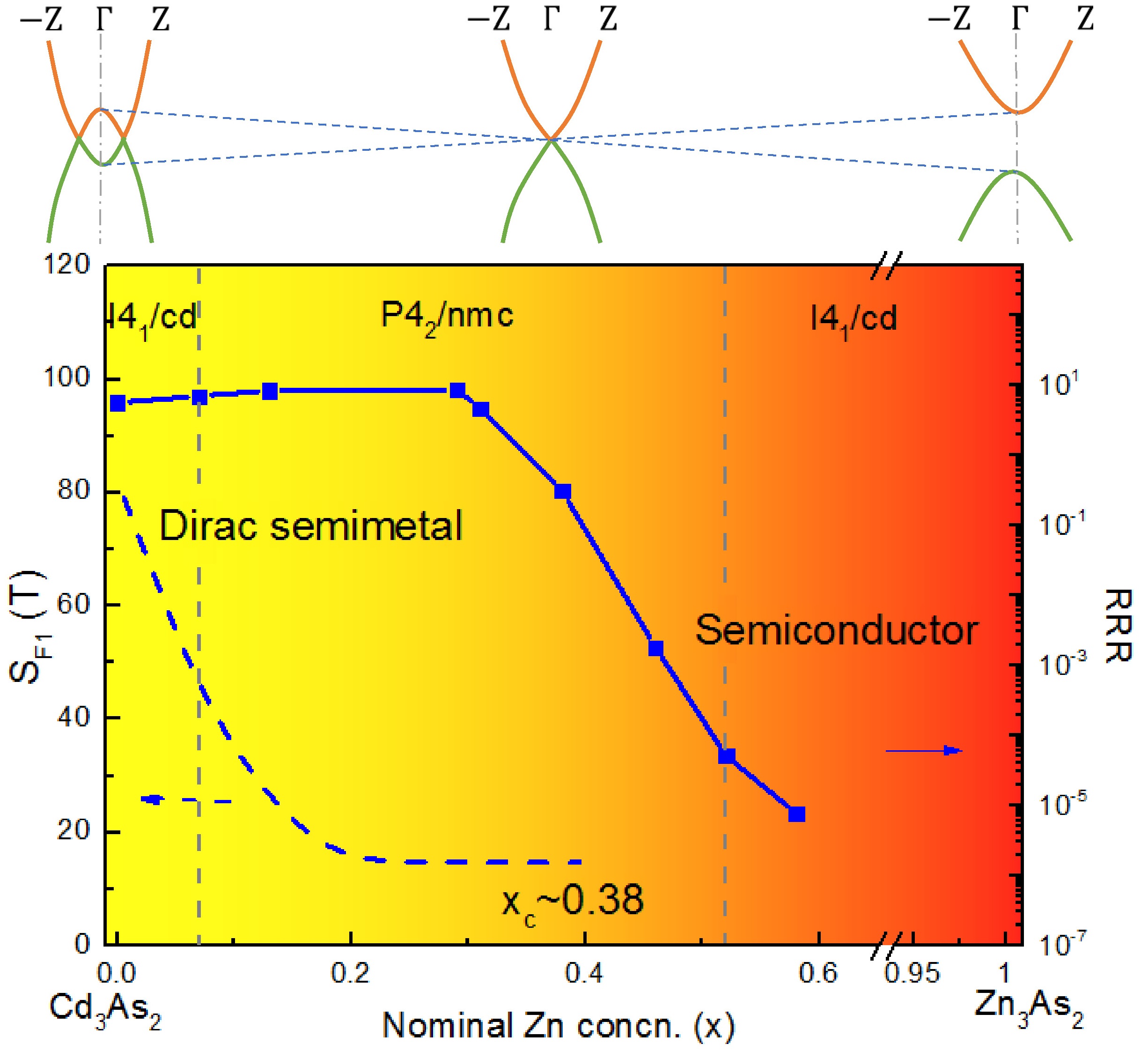}
	\caption{\textbf{Phase diagram for \bm{ $(Cd_{1-x} Zn_x)_3 As_2$}.} With the concentration of $Zn$ increasing, the samples transform from a topological Dirac semimetal to a semiconductor. The upper sketches of the band structure illustrate this transition. The structure transforms from  $I4_1cd$ to $P4_2/nmc$ then back to $I4_1cd$ with vertical dashed lines serving as rough boundaries. The background color presents the gradual change of resistivity at $2K$ as $x$ increases (inset of Fig. 4(a)). The changes of $S_{F1}$ and $RRR$ are plotted in the diagram as well.}
\end{figure}
\clearpage

\newpage

\begin{table}
	\centering
	\caption{\textbf{Parameters of the tested samples of different zinc concentration \bm{$x$}.} The dashed entries mean quantities missing.}
	\includegraphics[width=6.5in]{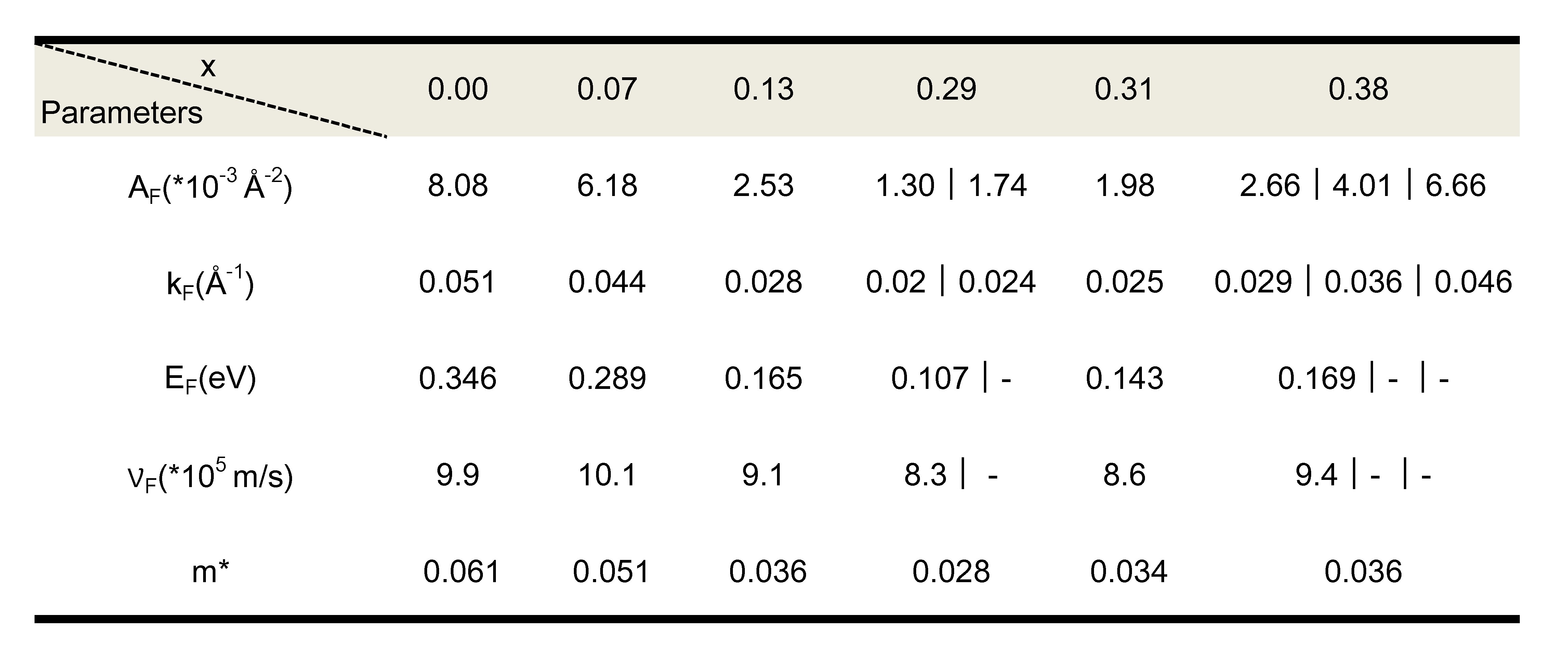}
	
\end{table}

\end{document}